\documentclass[eqsecnum,showpacs,preprint,amsmath,aps,nofootinbib]{revtex4}  %
\usepackage{epsfig}
\usepackage{amssymb}
\usepackage[dvips]{color}
\usepackage[latin1]{inputenc}
\usepackage{graphicx}

\def\beq{\begin{equation}}
\def\eeq{\end{equation}}
\def\beqq{\begin{eqnarray}}
\def\eeqq{\end{eqnarray}}

\newcommand{\bdm}{\begin{displaymath}}
\newcommand{\edm}{\end{displaymath}}

\def\pmb#1{\setbox0=\hbox{$#1$}%
  \kern-.025em\copy0\kern-\wd0
  \kern.05em\copy0\kern-\wd0
  \kern-.025em\raise.0433em\box0}

\makeatletter
\renewcommand*{\@fnsymbol}[1]{\ensuremath{\ifcase#1\or *\or \dagger\or
    \ddagger\or 
   \mathsection\or **\or \dagger\dagger
   \or \ddagger\ddagger \else\@ctrerr\fi}}
\makeatother

\begin{document}
\title{Full three-body problem in effective-field-theory models
of gravity}

\author{Emmanuele Battista}
\email[E-mail: ]{ebattista@na.infn.it}
\affiliation{Dipartimento di Fisica, Complesso Universitario 
di Monte S. Angelo, Via Cintia Edificio 6, 80126 Napoli, Italy\\
Istituto Nazionale di Fisica Nucleare, Sezione di Napoli, Complesso Universitario di Monte
S. Angelo, Via Cintia Edificio 6, 80126 Napoli, Italy}

\author{Giampiero Esposito}
\email[E-mail: ]{gesposit@na.infn.it}
\affiliation{Istituto Nazionale di Fisica Nucleare, Sezione di
Napoli, Complesso Universitario di Monte S. Angelo, 
Via Cintia Edificio 6, 80126 Napoli, Italy}

\date{\today}

\begin{abstract}
Recent work in the literature has studied the restricted 
three-body problem within the framework of effective-field-theory
models of gravity. This paper extends such a program by considering
the full three-body problem, when the Newtonian potential is replaced
by a more general central potential which depends on the mutual
separations of the three bodies. The general form of the equations
of motion is written down, and they are studied 
when the interaction potential reduces to the quantum-corrected
central potential considered recently in the literature.
A recursive algorithm is found for solving the associated variational equations,
which describe small departures from given periodic solutions of the equations of motion.
Our scheme involves repeated application of a $2 \times 2$ matrix of first-order
linear differential operators.
\end{abstract}


\maketitle

\section{Introduction}

As was stressed by Poincar\'e in his landmark work on the (restricted)
three-body problem \cite{P1890}, the main aim of celestial mechanics is 
not the one of evaluating the astronomical ephemeris, but rather to ascertain
whether Newtonian theory remains the most appropriate tool for 
investigating celestial gravity \cite{P1892}, at least 
(we would say) within the solar system. 
With hindsight, this statement is not completely superseded 
by current developments in gravitational theories, 
provided in its formulation one replaces Newtonian theory
by Einstein's general relativity, which has been challenged
over the years by several competing theories (e.g.
Brans-Dicke, $f(R)$, ...), to be tested both in the solar system and on
extra-galactic scales. In particular, the hybrid scheme where the
Newtonian potential receives classical and quantum corrections from the
calculational recipes of effective field theories has been studied in
detail in Refs. \cite{D94,D94b,D94c,MV95,HL95,ABS97,KK02,D03}
and has been applied recently to the
investigation of the restricted three-body problem of celestial
mechanics \cite{BE14}. Interestingly, we have found that the 
consideration of this problem makes it possible to discriminate
competing models of quantum corrections to the Newtonian potential, 
and that the evaluation of first-order stability is, at least in
principle, slightly affected by such tiny corrections, because the
planetoid is no longer at equal distance from the two bodies of
large mass, although the expected displacement from the classical
equilateral triangle \cite{Pars65} picture is very small and not so obviously
observable (see Appendix A).

\begin{figure}
\includegraphics[scale=0.35]{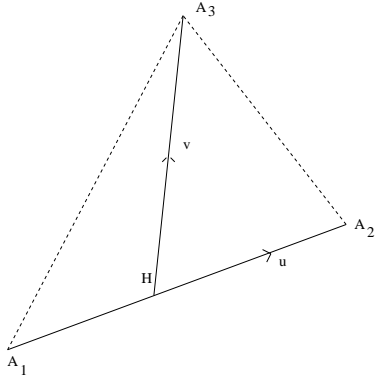}
\caption{The $3$ bodies $A_{1},A_{2},A_{3}$, the center of mass $H$ of $A_{1}$ and
$A_{2}$, the vector $\vec u$ joining $A_{1}$ to $A_{2}$, and the vector $\vec v$
joining $H$ to $A_{3}$ \cite{Pars65}.}
\end{figure}

It has been therefore our aim to go one step further by assessing from
this novel perspective the full three-body problem of celestial mechanics.
In this setting, three bodies $A_{1},A_{2},A_{3}$ having masses 
$m_{1},m_{2},m_{3}$, respectively, move in space under the action of
their mutual gravitational attraction, whose functional form remains
unspecified for the time being (see Secs. II and III). Following
Ref. \cite{Pars65} we take fixed rectangular axes and denote the
coordinates of $A_{r}$ at time $t$ by $x_{r},y_{r},z_{r}$. The
coordinates of the center or mass $D$ of the three bodies are instead
denoted by block capital letters $X,Y,Z$, so that, on denoting
by $M \equiv m_{1}+m_{2}+m_{3}$ the mass of the whole system, one can write
\begin{equation}
MX=\sum_{r=1}^{3}m_{r}x_{r}, \; MY=\sum_{r=1}^{3}m_{r}y_{r}, \;
MZ=\sum_{r=1}^{3}m_{r}z_{r}.
\label{(1.1)}
\end{equation}
Let the vector ${\overrightarrow {A_{1}A_{2}}}$ be ${\vec u}$, and let the vector
${\overrightarrow {HA_{3}}}$ ($H$ being the center of mass of $A_{1}$ and 
$A_{2}$) be ${\vec v}$ (Fig. 1). Thus, by defining the parameters
\begin{equation}
\alpha_{1} \equiv {m_{1}\over (m_{1}+m_{2})}, \;
\alpha_{2} \equiv 1-\alpha_{1},
\label{(1.2)}
\end{equation}
the vector ${\overrightarrow {A_{2}A_{3}}}$ is $(-\alpha_{1}{\vec u}+{\vec v})$,
while the vector ${\overrightarrow {A_{1}A_{3}}}$ is
$(\alpha_{2}{\vec u}+{\vec v})$. Hereafter, we denote by $(x,y,z)$ the
components of ${\vec u}$, and by $(\xi,\eta,\zeta)$ the components
of ${\vec v}$. The positions and velocities of the three bodies at
$t=0$ are prescribed, and the problem is to determine their position
at any subsequent time.

Section II builds the Lagrangian and arrives at the general form
of the equations of motion. Section III considers the choice of quantum  
corrected potential. Section IV writes such a general set
of equations when the potential $U$ takes precisely the form considered
in our previous paper \cite{BE14} and suggested by the work in Refs.
\cite{D94,D94b,MV95,HL95,ABS97,KK02,D03}. Variational equations 
are investigated in Sec. V, and a general solution algorithm of variational equations
is derived in Sec. VI. Concluding remarks and open problems are presented in Sec. VII.

\section{Lagrangian and equations of motion}

With the coordinates introduced at the end of the Introduction, the
kinetic energy $T$ can be expressed by means of the relation
\cite{Pars65}
\begin{equation}
T={M\over 2}({\dot X}^{2}+{\dot Y}^{2}+{\dot Z}^{2})
+{1\over 2}\sum_{r < s}{m_{r}m_{s}\over M}v_{rs}^{2},
\label{(2.1)}
\end{equation}
where $v_{rs}$ is the speed of $A_{s}$ relative to $A_{r}$, i.e.
\begin{equation}
v_{rs}^{2}=({\dot x}_{s}-{\dot x}_{r})^{2}
+({\dot y}_{s}-{\dot y}_{r})^{2}
+({\dot z}_{s}-{\dot z}_{r})^{2}.
\label{(2.2)}
\end{equation}
On defining the ``reduced masses''
\begin{equation}
m \equiv {m_{1}m_{2}\over (m_{1}+m_{2})}, \;
\mu \equiv {(m_{1}+m_{2})m_{3}\over (m_{1}+m_{2}+m_{3})},
\label{(2.3)}
\end{equation}
the $x$-terms in $T$ arising from the motion relative to $D$
give \cite{Pars65}
\begin{equation}
{1\over 2M}\Bigr[m_{2}m_{3}(-\alpha_{1}{\dot x}
+{\dot \xi})^{2}+m_{3}m_{1}(\alpha_{2}{\dot x}+{\dot \xi})^{2}
+m_{1}m_{2}{\dot x}^{2}\Bigr] 
={m\over 2}{\dot x}^{2}+{\mu \over 2}{\dot \xi}^{2}.
\label{(2.4)}
\end{equation}
One has now to add the corresponding formulas for $y$ and $z$,
which yields the neat result
\begin{equation}
T={M \over 2}\Bigr({\dot X}^{2}+{\dot Y}^{2}+{\dot Z}^{2}\Bigr)
+{m\over 2}({\dot x}^{2}+{\dot y}^{2}+{\dot z}^{2})
+{\mu \over 2}({\dot \xi}^{2}+{\dot \eta}^{2}+{\dot \zeta}^{2}).
\label{(2.5)}
\end{equation}

In Newtonian theory, one proceeds by assuming at this stage a 
potential of the form \cite{Pars65}
\begin{equation}
U=G \left({m_{2}m_{3}\over r_{1}}+{m_{3}m_{1}\over r_{2}}
+{m_{1}m_{2}\over r_{3}}\right),
\label{(2.6)}
\end{equation}
having defined
\begin{equation}
(r_{1})^{2} \equiv (-\alpha_{1}{\vec u}+{\vec v}) \cdot
(-\alpha_{1}{\vec u}+{\vec v})
=(-\alpha_{1}x+\xi)^{2}+(-\alpha_{1}y+\eta)^{2}
+(-\alpha_{1}z+\zeta)^{2},
\label{(2.7)}
\end{equation}
\begin{equation}
(r_{2})^{2} \equiv (\alpha_{2}{\vec u}+{\vec v}) \cdot
(\alpha_{2}{\vec u}+{\vec v})
=(\alpha_{2}x+\xi)^{2}+(\alpha_{2}y+\eta)^{2}
+(\alpha_{2}z+\zeta)^{2},
\label{(2.8)}
\end{equation}
\begin{equation}
(r_{3})^{2} \equiv {\vec u} \cdot {\vec u}
=x^{2}+y^{2}+z^{2}.
\label{(2.9)}
\end{equation}
In our case, although we keep using the classical concepts of
kinetic energy and center of mass, we depart from classical
Newtonian theory by assuming that $U$ can be a more general
function of $r_{1},r_{2},r_{3}$, i.e.
\begin{equation}
U=U(r_{1},r_{2},r_{3})=\sum_{k=1}^{3} U_{k}(r_{k}).
\label{(2.10)}
\end{equation}
We will first derive the equations of motion resulting from the 
general choice (2.10), and we will eventually look for explicit
solutions with a choice of $U$ inspired by Refs.
\cite{D94,D94b,D03,BE14}.

By virtue of (\ref{(2.5)}) and (\ref{(2.10)}), the
Lagrangian equations of motion read as
\begin{equation}
M {\ddot X}={\partial U \over \partial X}, \;
m {\ddot x}={\partial U \over \partial x}, \;
\mu {\ddot \xi}={\partial U \over \partial \xi},
\label{(2.11)}
\end{equation}
supplemented by the corresponding second-order equations for
$(Y,y,\eta)$ and $(Z,z,\zeta)$. Since, from (\ref{(2.10)}), 
$U$ is independent of $X,Y,Z$, one has
\begin{equation}
{\ddot X}={\ddot Y}={\ddot Z}=0,
\label{(2.12)}
\end{equation}
which means that the center of mass $D$ moves uniformly in a
straight line. We may even assume that $D$ remains at rest without
losing generality, and the remaining Eqs. for $m \ddot x$ and
$\mu \ddot \xi$ in (\ref{(2.11)}) can be obtained by setting
\begin{equation}
U_{,r_{j}} \equiv {\partial U \over \partial r_{j}}, 
\; \; \; \forall j=1,2,3,
\label{(2.13)}
\end{equation}
and writing patiently the partial derivatives
\begin{equation}
{\partial U \over \partial x}=U_{,r_{1}}
{\partial r_{1}\over \partial x}
+U_{,r_{2}}{\partial r_{2}\over \partial x} 
+U_{,r_{3}}{\partial r_{3}\over \partial x},
\label{(2.14)}
\end{equation}
\begin{equation}
{\partial U \over \partial \xi}=U_{,r_{1}}
{\partial r_{1}\over \partial \xi}
+U_{,r_{2}}{\partial r_{2}\over \partial \xi}.
\label{(2.15)}
\end{equation}
In light of (\ref{(2.7)})-(\ref{(2.9)}), one arrives therefore
at the formulas
\begin{equation}
{\partial U \over \partial x}=-Ax+B \xi ,
\label{(2.16)}
\end{equation}
\begin{equation}
{\partial U \over \partial \xi}=Bx-C \xi,
\label{(2.17)}
\end{equation}
where we have defined (cf. Eq. (29.10.11) in Ref. \cite{Pars65})
\begin{equation}
A \equiv -{\alpha_{1}^{2}\over r_{1}}U_{,r_{1}}
-{\alpha_{2}^{2}\over r_{2}}U_{,r_{2}}
-{1\over r_{3}}U_{,r_{3}},
\label{(2.18)}
\end{equation}
\begin{equation}
B \equiv {\alpha_{2}\over r_{2}}U_{,r_{2}}
-{\alpha_{1}\over r_{1}}U_{,r_{1}},
\label{(2.19)}
\end{equation}
\begin{equation}
C \equiv -{1\over r_{1}}U_{,r_{1}}-{1\over r_{2}}U_{,r_{2}}.
\label{(2.20)}
\end{equation}
After writing the corresponding equations for $(y,\eta)$ and
$(z,\zeta)$ one obtains eventually, bearing in mind that 
${\vec u}$ has components $(x,y,z)$, while ${\vec v}$ has
components $(\xi,\eta,\zeta)$, the equations of motion in matrix form
\begin{equation}
\left(\begin{matrix}
m {{\rm d}^{2}\over {\rm d}t^{2}}+A & -B \cr
-B & \mu {{\rm d}^{2}\over {\rm d}t^{2}}+C
\end{matrix}
\right)
\left(\begin{matrix}
{\vec u} \cr {\vec v} 
\end{matrix}\right)=0.
\label{(2.21)}
\end{equation}
Such a scheme tells us that the full three-body problem is equivalent
to a system of two particles, i.e. a particle of mass $m$ at 
$(x,y,z)$ and a particle of mass $\mu$ at $(\xi,\eta,\zeta)$.

The integrals of angular momentum are found to take 
the form \cite{Pars65}
\begin{equation}
M(Y {\dot Z}-Z{\dot Y})+m(y{\dot z}-z{\dot y})
+\mu (\eta {\dot \zeta}-\zeta {\dot \eta})=a,
\label{(2.22)}
\end{equation} 
\begin{equation}
M(Z {\dot X}-X{\dot Z})+m(z{\dot x}-x{\dot z})
+\mu (\zeta {\dot \xi}-\xi {\dot \zeta})=b,
\label{(2.23)}
\end{equation} 
\begin{equation}
M(X {\dot Y}-Y{\dot X})+m(x{\dot y}-y{\dot x})
+\mu (\xi {\dot \eta}-\eta {\dot \xi})=c.
\label{(2.24)}
\end{equation} 
Since the center of mass moves uniformly in a straight line, the
terms $M(Y{\dot Z}-Z {\dot Y})$ and
$[m(y{\dot z}-z{\dot y})+\mu (\eta {\dot \zeta}
-\zeta{\dot \eta})]$ in (\ref{(2.22)}) are separately constant,
and similarly in Eqs. (\ref{(2.23)}) and (\ref{(2.24)}).
Indeed, one finds from Eq. (\ref{(2.11)})
\begin{equation}
{{\rm d}\over {\rm d}t}[m(y{\dot z}-z{\dot y})
+\mu(\eta {\dot \zeta}-\zeta{\dot \eta})]
=\left(y{\partial \over \partial z}-z {\partial \over \partial y}
+\eta {\partial \over \partial \zeta}
-\zeta {\partial \over \partial \eta}\right)U,
\label{(2.25)}
\end{equation}
which vanishes, because $U$ depends on $r_{1},r_{2},r_{3}$ 
separately, according to Eq. (\ref{(2.10)}), and the following
identity holds:
\begin{equation}
\left(y{\partial \over \partial z}
-z{\partial \over \partial y}
+\eta {\partial \over \partial \zeta}
-\zeta {\partial \over \partial \eta}\right)r_{k}=0, \;
\forall k=1,2,3.
\label{(2.26)}
\end{equation}
The forces are not in the line joining the particles, but 
their moment about the origin is
$$
{\vec u} \times (-A {\vec u}+B {\vec v})
+{\vec v} \times (B {\vec u}-C {\vec v}),
$$
which vanishes by virtue of the skew-symmetry of the vector 
product. Hence the angular momentum about the origin remains
constant as in Newtonian theory \cite{Pars65}.

\section{A choice of quantum corrected potential}

After having written the equations of motion in a rather general
form, we cannot attempt any integration without an explicit form
of the potential function. For this purpose, we now investigate
the implications of assuming that the classical potential (2.6) can be
replaced by a quantum corrected potential according to the recipes
considered in Ref. \cite{BE14}. This means that the general formula
(2.10) can take the form
\begin{eqnarray}
U(r_{1},r_{2},r_{3})&=& 
{G m_{2}m_{3}\over r_{1}}
\left(1+\kappa_{23} {G \over c^{2}}
{(m_{2}+m_{3})\over r_{1}}
+\kappa {l_{P}^{2} \over (r_{1})^{2}}\right)
\nonumber \\
&+&{G m_{1}m_{3}\over r_{2}}
\left(1+\kappa_{13}  {G \over c^{2}}
{(m_{1}+m_{3})\over r_{2}}
+\kappa {l_{P}^{2} \over (r_{2})^{2}}\right)
\nonumber \\
&+&{G m_{1}m_{2}\over r_{3}}
\left(1+\kappa_{12}  {G \over c^{2}}
{(m_{1}+m_{2})\over r_{3}}
+\kappa {l_{P}^{2} \over (r_{3})^{2}}\right),
\label{(3.1)}
\end{eqnarray}
where the parameters $\kappa,\kappa_{12},\kappa_{23}$ and $\kappa_{13}$
are dimensionless, and $l_{P}$ is the Planck length. 
We stress that $\kappa_{23},\kappa_{13},\kappa_{12}$ depend on $\kappa$
because they are part of a calculational recipe that yields, at the
same time, a post-Newtonian term and a fully quantum term. We are not
evaluating the quantum corrections to relativistic celestial mechanics. 
By using Fourier-transform techniques, the ${1\over q^{2}}$ term in momentum space
leads to ${1\over r}$, while ${1\over q^{2}} \times \sqrt{q^{2}}$ and 
${1\over q^{2}} \times q^{2} \log(q^{2})$ lead to ${1\over r^{2}}$ and
${1\over r^{3}}$, respectively. The corrections obtained in Ref. \cite{D03}
result from all one-loop diagrams that can contribute to the scattering of two
masses. One then finds nonanalytic corrections of the form
$Gm \sqrt{q^{2}}$ and $Gq^{2}\log(q^{2})$, as well as analytic terms $Gq^{2}$.

The first derivatives of such a potential, to be used in the definitions
(2.18)-(2.20) of the functions $A,B,C$ read therefore as
\begin{equation}
U_{,r_{1}}=-{Gm_{2}m_{3}\over (r_{1})^{2}}
\left(1+2 \kappa_{23}  {G \over c^{2}}
{(m_{2}+m_{3})\over r_{1}}
+3 \kappa {l_{P}^{2}\over (r_{1})^{2}}\right),
\label{(3.2)}
\end{equation}
\begin{equation}
U_{,r_{2}}=-{Gm_{1}m_{3}\over (r_{2})^{2}}
\left(1+2 \kappa_{13}  {G \over c^{2}}
{(m_{1}+m_{3})\over r_{2}}
+3 \kappa {l_{P}^{2}\over (r_{2})^{2}}\right),
\label{(3.3)}
\end{equation}
\begin{equation}
U_{,r_{3}}=-{Gm_{1}m_{2}\over (r_{3})^{2}}
\left(1+2 \kappa_{12}  {G \over c^{2}}
{(m_{1}+m_{2})\over r_{3}}
+3 \kappa {l_{P}^{2}\over (r_{3})^{2}}\right).
\label{(3.4)}
\end{equation}

\section{Hamiltonian equations of motion}

The equations of motion (2.21) are Lagrangian second-order equations 
of motion. They can be re-expressed as a coupled set of twelve 
first-order Hamiltonian equations as follows:
\begin{equation}
{{\rm d} \over {\rm d}t}x=p_{x},
\label{(4.1)}
\end{equation}
\begin{equation}
{{\rm d} \over {\rm d}t}y=p_{y},
\label{(4.2)}
\end{equation}
\begin{equation}
{{\rm d} \over {\rm d}t}z=p_{z},
\label{(4.3)}
\end{equation}
\begin{equation}
{{\rm d} \over {\rm d}t}\xi=p_{\xi},
\label{(4.4)}
\end{equation}
\begin{equation}
{{\rm d} \over {\rm d}t}\eta=p_{\eta},
\label{(4.5)}
\end{equation}
\begin{equation}
{{\rm d} \over {\rm d}t}\zeta=p_{\zeta},
\label{(4.6)}
\end{equation}
\begin{equation}
{{\rm d}\over {\rm d}t}p_{x}=-{1\over m}(Ax-B \xi),
\label{(4.7)}
\end{equation}
\begin{equation}
{{\rm d}\over {\rm d}t}p_{y}=-{1\over m}(Ay-B \eta),
\label{(4.8)}
\end{equation}
\begin{equation}
{{\rm d}\over {\rm d}t}p_{z}=-{1\over m}(Az-B \zeta),
\label{(4.9)}
\end{equation}
\begin{equation}
{{\rm d}\over {\rm d}t}p_{\xi}=-{1\over \mu}(C \xi-Bx),
\label{(4.10)}
\end{equation}
\begin{equation}
{{\rm d}\over {\rm d}t}p_{\eta}=-{1\over \mu}(C \eta-By),
\label{(4.11)}
\end{equation}
\begin{equation}
{{\rm d}\over {\rm d}t}p_{\zeta}=-{1\over \mu}(C \zeta-Bz).
\label{(4.12)}
\end{equation}
We need therefore twelve initial conditions to integrate these 
equations of motion. Hereafter it is convenient to introduce the
$6$-tuple of position variables
\begin{equation}
x_{i} \equiv (x,y,z,\xi,\eta,\zeta) 
\equiv (x_{1},...,x_{6}),
\label{(4.13)}
\end{equation}
and the $6$-tuple of momentum variables
\begin{equation}
y_{i} \equiv (p_{x},p_{y},p_{z},p_{\xi},p_{\eta},p_{\zeta})
\equiv (p_{1},...,p_{6}).
\label{(4.14)}
\end{equation}
The equations (4.1)-(4.12) can be therefore further re-expressed 
in the canonical form \cite{P1890,P1892}
\begin{equation}
{{\rm d}\over {\rm d}t}x_{i}={\partial F \over \partial y_{i}},
\; {{\rm d}\over {\rm d}t}y_{i}=-{\partial F \over \partial x_{i}},
\label{(4.15)}
\end{equation}
where the function $F$ is given by
\begin{equation}
F(x_{1},...,x_{6},y_{1},...,y_{6})=\sum_{i=1}^{6}
{y_{i}^{2}\over 2}+f(x_{1},...,x_{6}),
\label{(4.16)}
\end{equation}
and $f$ solves the system of partial differential equations 
obtainable from (4.7)-(4.12), i.e.,
\begin{equation}
{\partial f \over \partial x}={1\over m}(Ax-B\xi),
\label{(4.17)}
\end{equation}
\begin{equation}
{\partial f \over \partial y}={1\over m}(Ay-B\eta),
\label{(4.18)}
\end{equation}
\begin{equation}
{\partial f \over \partial z}={1\over m}(Az-B\zeta),
\label{(4.19)}
\end{equation}
\begin{equation}
{\partial f \over \partial \xi}={1\over \mu}(C \xi-Bx),
\label{(4.20)}
\end{equation}
\begin{equation}
{\partial f \over \partial \eta}={1\over \mu}(C \eta-By),
\label{(4.21)}
\end{equation}
\begin{equation}
{\partial f \over \partial \zeta}={1\over \mu}(C \zeta-Bz),
\label{(4.22)}
\end{equation}
the functions $A(x_{1},...,x_{6}),B(x_{1},...,x_{6}),
C(x_{1},...,x_{6})$ being defined by (2.18)-(2.20), supplemented by
(2.7)-(2.9) and (3.2)-(3.4).

At this stage we can exploit a fundamental theorem proved by
Poincar\'e \cite{P1890,P1892}, according to which, {\it if the
equations (4.15), which depend on a parameter $\rho$, possess
for $\rho=0$ a periodic solution whose characteristic exponents}
(see Appendix) {\it are all nonvanishing, they have again a periodic
solution for small values of $\rho$}. In our case, the small
parameter $\rho$ is the Planck length $l_{P}$, 
and when $\rho=0$ we revert
to the three-body problem in post-Newtonian mechanics, for which,
in the circular restricted case, one knows from recent work
\cite{HuangWu} that orbits may be unstable, or bounded chaotic, or bounded regular. 
In the case of Newtonian mechanics instead,  
Chenciner and Montgomery \cite{Chenciner00} have found
a class of solutions where three bodies of equal mass move periodically
on the plane along the same curve. The periodic orbit has zero
angular momentum, and the three bodies chase each other around a fixed
eight-shaped curve. Such an orbit visits in turn every Euler
configuration in which one of the bodies sits at the midpoint of the
segment defined by the other two.

To sum up, we have found that, by virtue of the Poincar\'e
theorem on periodic solutions and of the extreme smallness of the Planck
length, also our quantum corrected potential (3.1) may lead to 
periodic solutions. This is a novel perspective on a smooth matching
between classical and quantum-corrected three-body problems.

\section{Variational equations}

Following Ref. \cite{P1890}, let us now revert to the Eqs. (4.15),
and let us assume that a periodic solution has been found
\begin{equation}
x_{i}=\varphi_{i}(t), \; y_{i}=\psi_{i}(t).
\label{(5.1)}
\end{equation}
With the notation in Appendix B, we now investigate an algorithm
for the evaluation of characteristic exponents. For this purpose,
we consider small disturbances of such periodic solutions, written
in the form
\begin{equation}
{\tilde x}_{i}=\varphi_{i}(t)+\xi_{i}, \;
{\tilde y}_{i}=\psi_{i}(t)+\eta_{i},
\label{(5.2)}
\end{equation}
and we form the variational equations (cf. Eq. (B4)) 
resulting from the linearized approximation, i.e.
\begin{equation}
{{\rm d}\over {\rm d}t}\xi_{i}
=\sum_{k=1}^{6}\Bigr[F_{,y_{i}x_{k}}\xi_{k}
+F_{,y_{i}y_{k}}\eta_{k}\Bigr],
\label{(5.3)}
\end{equation}
\begin{equation}
{{\rm d}\eta_{i}\over {\rm d}t}
=-\sum_{k=1}^{6}\Bigr[F_{,x_{i}x_{k}}\xi_{k}
+F_{,x_{i}y_{k}}\eta_{k}\Bigr],
\label{(5.4)}
\end{equation}
where a subscript consisting of a comma followed by a variable
denotes partial derivative with respect to that variable,
e.g. $F_{,x_{k}} \equiv {\partial F \over \partial x_{k}}$.
Following Ref. \cite{P1890}, we try to integrate these variational
equations by setting
\begin{equation}
\xi_{i}={\rm e}^{\alpha t}S_{i}, \;
\eta_{i}={\rm e}^{\alpha t}T_{i},
\label{(5.5)}
\end{equation}
$S_{i}$ and $T_{i}$ being unknown periodic functions of $t$. The work in
Ref. \cite{P1890} provided a remarkable proof that if, when 
$\rho=0$, the characteristic exponents are vanishing,
then for small but nonvanishing values of $\rho$ one can expand
$\alpha,S_{i}$ and $T_{i}$ in the form
\begin{equation}
\alpha \sim \sum_{j=1}^{N}\alpha_{j}\rho^{{j \over 2}},
\label{(5.6)}
\end{equation}
\begin{equation}
S_{i} \sim \sum_{l=0}^{N}S_{i}^{l}\rho^{{l \over 2}},
\label{(5.7)}
\end{equation}
\begin{equation}
T_{i} \sim \sum_{l=0}^{N}T_{i}^{l}\rho^{{l \over 2}}.
\label{(5.8)}
\end{equation}
{\it This framework is complementary to the one mentioned at the end
of Sec. IV}, where we mentioned the Poincar\'e theorem on the persistence
of periodic solutions at small $\rho$. That theorem assumes instead
that, at $\rho=0$, the characteristic exponents $\alpha$ are all
nonvanishing. 
   
We now insert the formulas (5.5)--(5.8) into the variational 
equations (5.3) and (5.4), assuming for $F$ the asymptotic expansion
(in Sec. VI we will see that $F_{1}$ and ${\rm O}(\rho^{3})$ vanish identically
in our model)
\begin{equation}
F \sim F_{0}+\rho F_{1}+\rho^{2}F_{2}+{\rm O}(\rho^{3}).
\label{(5.9)}
\end{equation}
Now the asymptotic expansion of left-hand side of 
the variational equations (5.3) and (5.4) yields
\begin{equation}
{{\rm d}\over {\rm d}t}\xi_{i} \sim {\rm e}^{\alpha t}
\left[{{\rm d}S_{i}^{0}\over {\rm d}t}
+\left(\alpha_{1}S_{i}^{0}+{{\rm d}S_{i}^{1}\over {\rm d}t}
\right)\sqrt{\rho}
+\left(\alpha_{1}S_{i}^{1}+\alpha_{2}S_{i}^{0}
+{{\rm d}S_{i}^{2}\over {\rm d}t}\right)\rho
+{\rm O}(\rho^{{3\over 2}})\right],
\label{(5.10)}
\end{equation}
\begin{equation}
{{\rm d}\over {\rm d}t}\eta_{i} \sim {\rm e}^{\alpha t}
\left[{{\rm d}T_{i}^{0}\over {\rm d}t}
+\left(\alpha_{1}T_{i}^{0}+{{\rm d}T_{i}^{1}\over {\rm d}t}
\right)\sqrt{\rho}
+\left(\alpha_{1}T_{i}^{1}+\alpha_{2}T_{i}^{0}
+{{\rm d}T_{i}^{2}\over {\rm d}t}\right)\rho
+{\rm O}(\rho^{{3\over 2}})\right],
\label{(5.11)}
\end{equation}
so that comparison of coefficients of equal powers of $\rho$ yields for all $i=1,...,6$,
up to first order in $\rho$, the equations
\begin{equation}
{{\rm d}S_{i}^{0}\over {\rm d}t}=\sum_{k=1}^{6}
\Bigr({F_{0}}_{,y_{i}x_{k}}S_{k}^{0}
+{F_{0}}_{,y_{i}y_{k}}T_{k}^{0}\Bigr),
\label{(5.12)}
\end{equation}
\begin{equation}
\alpha_{1}S_{i}^{0}+{{\rm d}S_{i}^{1}\over {\rm d}t}
=\sum_{k=1}^{6}\Bigr({F_{0}}_{,y_{i}x_{k}}S_{k}^{1}
+{F_{0}}_{,y_{i}y_{k}}T_{k}^{1}\Bigr),
\label{(5.13)}
\end{equation}
\begin{equation}
\alpha_{1}S_{i}^{1}+\alpha_{2}S_{i}^{0}
+{{\rm d}S_{i}^{2}\over {\rm d}t}
=\sum_{k=1}^{6}\Bigr({F_{0}}_{,y_{i}x_{k}}S_{k}^{2}
+{F_{1}}_{,y_{i}x_{k}}S_{k}^{0}
+{F_{0}}_{,y_{i}y_{k}}T_{k}^{2}
+{F_{1}}_{,y_{i}y_{k}}T_{k}^{0}\Bigr),
\label{(5.14)}
\end{equation}
\begin{equation}
{{\rm d}T_{i}^{0}\over {\rm d}t}=-\sum_{k=1}^{6}
\Bigr({F_{0}}_{,x_{i}x_{k}}S_{k}^{0}
+{F_{0}}_{,x_{i}y_{k}}T_{k}^{0}\Bigr),
\label{(5.15)}
\end{equation}
\begin{equation}
\alpha_{1}T_{i}^{0}+{{\rm d}T_{i}^{1}\over {\rm d}t}
=-\sum_{k=1}^{6}\Bigr({F_{0}}_{,x_{i}x_{k}}S_{k}^{1}
+{F_{0}}_{,x_{i}y_{k}}T_{k}^{1}\Bigr),
\label{(5.16)}
\end{equation}
\begin{equation}
\alpha_{1}T_{i}^{1}+\alpha_{2}T_{i}^{0}
+{{\rm d}T_{i}^{2}\over {\rm d}t}
=-\sum_{k=1}^{6}\Bigr({F_{0}}_{,x_{i}x_{k}}S_{k}^{2}
+{F_{1}}_{,x_{i}x_{k}}S_{k}^{0}
+{F_{0}}_{,x_{i}y_{k}}T_{k}^{2}
+{F_{1}}_{,x_{i}y_{k}}T_{k}^{0}\Bigr).
\label{(5.17)}
\end{equation}
To begin, one should solve Eqs. (5.12) and (5.15) for 
$S_{i}^{0}$ and $T_{i}^{0}$, and insert them into (5.13) and
(5.16) to find $S_{i}^{1}$ and $T_{i}^{1}$, and iterate the
procedure to find $S_{i}^{2}$, $T_{i}^{2}$, ... as well as
$\alpha_{1},\alpha_{2},...$. 

\section{General solution of variational equations}

For the purpose of finding a general solution of variational equations, it may be helpful to elaborate
the equations of Sec. IV, where the potential term $U$ of Eq. (3.1) contains only a part of zeroth-order 
in $\rho \equiv l_{P}$ and a part of second order in $\rho$, and the same holds for the Hamiltonian
function $F$ in (4.16). More precisely, on defining
\begin{equation}
\gamma_{1}(r_{1}) \equiv -{G m_{2}m_{3}\over (r_{1})^{2}}
\left(1+2 \kappa_{23}{G \over c^{2}}{(m_{2}+m_{3})\over r_{1}}\right),
\label{(6.1)}
\end{equation}
\begin{equation}
\gamma_{2}(r_{2}) \equiv -{G m_{1}m_{3}\over (r_{2})^{2}}
\left(1+2 \kappa_{13}{G \over c^{2}}{(m_{1}+m_{3})\over r_{2}}\right),
\label{(6.2)}
\end{equation}
\begin{equation}
\gamma_{3}(r_{3}) \equiv -{G m_{1}m_{2}\over (r_{3})^{2}}
\left(1+2 \kappa_{12}{G \over c^{2}}{(m_{1}+m_{2})\over r_{3}}\right),
\label{(6.3)}
\end{equation}
we find that $A,B$ and $C$ in (2.18)-(2.20) take the form
\begin{equation}
A=A_{0}+\rho^{2}A_{2}, \; B=B_{0}+\rho^{2}B_{2}, C=C_{0}+\rho^{2}C_{2},
\label{(6.4)}
\end{equation}
where
\begin{equation}
A_{0}=-(\alpha_{1})^{2}{\gamma_{1}(r_{1})\over r_{1}}
-(\alpha_{2})^{2}{\gamma_{2}(r_{2})\over r_{2}}-{\gamma_{3}(r_{3})\over r_{3}},
\label{(6.5)}
\end{equation}
\begin{equation}
A_{2}=3G \kappa \left[(\alpha_{1})^{2}{m_{2}m_{3}\over (r_{1})^{5}}
+(\alpha_{2})^{2}{m_{1}m_{3}\over (r_{2})^{5}}
+{m_{1}m_{2}\over (r_{3})^{5}}\right],
\label{(6.6)}
\end{equation}
\begin{equation}
B_{0}=\alpha_{2}{\gamma_{2}(r_{2})\over r_{2}}
-\alpha_{1}{\gamma_{1}(r_{1})\over r_{1}},
\label{(6.7)}
\end{equation}
\begin{equation}
B_{2}=3G \kappa \left[\alpha_{1}{m_{2}m_{3}\over (r_{1})^{5}}
-\alpha_{2}{m_{1}m_{3}\over (r_{2})^{5}}\right],
\label{(6.8)}
\end{equation}
\begin{equation}
C_{0}=-{\gamma_{1}(r_{1})\over r_{1}}
-{\gamma_{2}(r_{2})\over r_{2}},
\label{(6.9)}
\end{equation}
\begin{equation}
C_{2}=3G \kappa \left[{m_{1}m_{3}\over (r_{2})^{5}}
+{m_{2}m_{3}\over (r_{1})^{5}}\right].
\label{(6.10)}
\end{equation}
At this stage, the coupled system (4.17)-(4.22) can be re-expressed in the form
\begin{equation}
{\partial f \over \partial x_{1}}={1\over m}(A_{0}x_{1}-B_{0}x_{4})
+{1 \over m}(A_{2}x_{1}-B_{2}x_{4})\rho^{2},
\label{(6.11)}
\end{equation}
\begin{equation}
{\partial f \over \partial x_{2}}={1\over m}(A_{0}x_{2}-B_{0}x_{5})
+{1 \over m}(A_{2}x_{2}-B_{2}x_{5})\rho^{2},
\label{(6.12)}
\end{equation}
\begin{equation}
{\partial f \over \partial x_{3}}={1\over m}(A_{0}x_{3}-B_{0}x_{6})
+{1 \over m}(A_{2}x_{3}-B_{2}x_{6})\rho^{2},
\label{(6.13)}
\end{equation}
\begin{equation}
{\partial f \over \partial x_{4}}={1\over \mu}(C_{0}x_{4}-B_{0}x_{1})
+{1 \over \mu}(C_{2}x_{4}-B_{2}x_{1})\rho^{2},
\label{(6.14)}
\end{equation}
\begin{equation}
{\partial f \over \partial x_{5}}={1\over \mu}(C_{0}x_{5}-B_{0}x_{2})
+{1 \over \mu}(C_{2}x_{5}-B_{2}x_{2})\rho^{2},
\label{(6.15)}
\end{equation}
\begin{equation}
{\partial f \over \partial x_{6}}={1\over \mu}(C_{0}x_{6}-B_{0}x_{3})
+{1 \over \mu}(C_{2}x_{6}-B_{2}x_{3})\rho^{2},
\label{(6.16)}
\end{equation}
where the left-hand sides can be further re-expressed upon writing
\begin{equation}
f(x_{1},...,x_{6})=f_{0}(x_{1},...,x_{6})+f_{2}(x_{1},...,x_{6})\rho^{2}.
\label{(6.17)}
\end{equation}
On the one hand, from (4.16) and (6.17) we have immediately that
\begin{equation}
F_{0,x_{i}y_{k}}=f_{0,x_{i}y_{k}}=0, \;
F_{0,y_{i}x_{k}}=(y_{i})_{,x_{k}}=0, \;
F_{0,y_{i}y_{k}}=\delta_{ik}, \;
F_{0,x_{i}x_{k}}=f_{0,x_{i}x_{k}}.
\label{(6.18)}
\end{equation}
On the other hand, from Eqs. (6.11)-(6.17), we find immediately the $6 \times 6$ matrix of partial
derivatives 
\begin{equation}
M_{ik}^{0} \equiv f_{0,x_{i}x_{k}},
\label{(6.19)}
\end{equation}
whose entries are written explicitly, for completeness, in Appendix C.
Now a patient application of (4.16), (6.17) and (6.18) to the Eqs. (5.12)-(5.17) reveals that,
for all $i=1,...,6$ (exploiting the vanishing of $F_{1}$ in our model)
\begin{equation}
\sum_{k=1}^{6}
\left(\begin{matrix}
\delta_{ik} {{\rm d}\over {\rm d}t} & -\delta_{ik} \cr
M_{ik}^{0} & \delta_{ik} {{\rm d}\over {\rm d}t}
\end{matrix}\right)
\left(\begin{matrix}
S_{k}^{0} \cr
T_{k}^{0}
\end{matrix}\right)=0,
\label{(6.20)}
\end{equation}
while, for higher-order terms, we find the inhomogeneous equations
\begin{equation}
\sum_{k=1}^{6}
\left(\begin{matrix}
\delta_{ik}{{\rm d}\over {\rm d}t} & -\delta_{ik} \cr
M_{ik}^{0} & \delta_{ik} {{\rm d}\over {\rm d}t}
\end{matrix}\right)
\left(\begin{matrix}
S_{k}^{n} \cr
T_{k}^{n}
\end{matrix}\right)
=-\sum_{l=0}^{n-1}\alpha_{n-l}
\left(\begin{matrix}
S_{i}^{l} \cr
T_{i}^{l}
\end{matrix}\right).
\label{(6.21)}
\end{equation}
For example, for the equations involving $\alpha_{1}$ and $\alpha_{2}$ we find
\begin{equation}
\sum_{k=1}^{6}
\left(\begin{matrix}
\delta_{ik}{{\rm d}\over {\rm d}t} & -\delta_{ik} \cr
M_{ik}^{0} & \delta_{ik} {{\rm d}\over {\rm d}t}
\end{matrix}\right)
\left(\begin{matrix}
S_{k}^{1} \cr
T_{k}^{1}
\end{matrix}\right)
=-\alpha_{1}
\left(\begin{matrix}
S_{i}^{0} \cr
T_{i}^{0}
\end{matrix}\right),
\label{(6.22)}
\end{equation}
\begin{equation}
\sum_{k=1}^{6}
\left(\begin{matrix}
\delta_{ik}{{\rm d}\over {\rm d}t} & -\delta_{ik} \cr
M_{ik}^{0} & \delta_{ik} {{\rm d}\over {\rm d}t}
\end{matrix}\right)
\left(\begin{matrix}
S_{k}^{2} \cr
T_{k}^{2}
\end{matrix}\right)
=-\alpha_{2}
\left(\begin{matrix}
S_{i}^{0} \cr
T_{i}^{0}
\end{matrix}\right)
-\alpha_{1}
\left(\begin{matrix}
S_{i}^{1} \cr
T_{i}^{1}
\end{matrix}\right).
\label{(6.23)}
\end{equation}

\subsection{The case when $\alpha$ does not vanish at $\rho=0$}

As we know from Sec. IV, it is at least equally important to study 
the case when the characteristic exponent does not vanish at
$\rho=0$ \cite{P1890,P1892}. In such a case, we assume that  
the asymptotic expansion (5.6) can be generalized by
adding the term $\alpha_{0}$, i.e.
\begin{equation}
\alpha \sim \sum_{l=0}^{N} \alpha_{l}\rho^{l \over 2}.
\label{(6.24)} 
\end{equation}
The method of Secs. V and VI leads eventually to equations that generalize (6.20)-(6.23) upon adding
$\alpha_{0}$ to the linear differential operator ${{\rm d}\over {\rm d}t}$, i.e.
\begin{equation}
\sum_{k=1}^{6}
\left(\begin{matrix}
\delta_{ik} \left({{\rm d}\over {\rm d}t}+\alpha_{0}\right) & -\delta_{ik} \cr
M_{ik}^{0} & \delta_{ik} \left({{\rm d}\over {\rm d}t}+\alpha_{0}\right)
\end{matrix}\right)
\left(\begin{matrix}
S_{k}^{0} \cr
T_{k}^{0}
\end{matrix}\right)=0,
\label{(6.25)}
\end{equation}
\begin{equation}
\sum_{k=1}^{6}
\left(\begin{matrix}
\delta_{ik} \left({{\rm d}\over {\rm d}t}+\alpha_{0}\right) & -\delta_{ik} \cr
M_{ik}^{0} & \delta_{ik} \left({{\rm d}\over {\rm d}t}+\alpha_{0}\right)
\end{matrix}\right)
\left(\begin{matrix}
S_{k}^{n} \cr
T_{k}^{n}
\end{matrix}\right)
=-\sum_{l=0}^{n-1}\alpha_{n-l}
\left(\begin{matrix}
S_{i}^{l} \cr
T_{i}^{l}
\end{matrix}\right).
\label{(6.26)}
\end{equation}

\subsection{Hamiltonian equations when $\rho=0$}

Our computational recipes are of little help unless we say what sort of periodic solutions we have in mind.
Since we are interested in small departures from classical theory, and $\rho \equiv l_{P}$ is the naturally
occurring parameter to describe such a scheme, we assume hereafter that the periodic solutions alluded to
in Eq. (5.1) are solutions of Eqs. (4.15) when $\rho=0$. With the notation in Eqs. (6.1)-(6.3), (6.5), (6.7)
and (6.9), the matrix (6.19) should be therefore evaluated along solutions of the coupled equations
\begin{equation}
{{\rm d}x_{i}\over {\rm d}t}=y_{i} \; \forall i=1,...,6 ,
\label{(6.27)}
\end{equation}
\begin{equation}
{{\rm d}y_{i}\over {\rm d}t}=-{1 \over m}(A_{0}x_{i}-B_{0}x_{i+3}) \; \forall i=1,2,3 ,
\label{(6.28)}
\end{equation}
\begin{equation}
{{\rm d}y_{i}\over {\rm d}t}=-{1 \over \mu}(C_{0}x_{i}-B_{0}x_{i-3}) \; \forall i=4,5,6 .
\label{(6.29)}
\end{equation}
The desired periodic solutions, whose existence is a special rather than generic property
\cite{P1890,P1892,HuangWu}, can be written in the form
\begin{equation}
x_{i}=\sum_{l=0}^{\infty}D_{il} \sin(\omega_{il}t+\varphi_{il}),
\label{(6.30)}
\end{equation}
\begin{equation}
y_{i}=\sum_{l=0}^{\infty}E_{il} \sin(\omega_{il}t+\gamma_{il}).
\label{(6.31)}
\end{equation}
When we insert such Fourier expansions into the system (6.27)-(6.29) we have to bear in mind that 
$A_{0},B_{0},C_{0}$ in (6.5), (6.7), (6.9) depend on $x_{1},...,x_{6}$ because Eqs. (2.7)-(2.9)
can be re-expressed in the form
\begin{equation}
(r_{1})^{2}=\sum_{k=1}^{3} (\alpha_{1}x_{k}-x_{k+3})^{2}, \;
(r_{2})^{2}=\sum_{k=1}^{3} (\alpha_{2}x_{k}+x_{k+3})^{2}, \;
(r_{3})^{2}=\sum_{k=1}^{3}(x_{k})^{2}.
\label{(6.32)}
\end{equation}

\section{Concluding remarks and open problems}

The equations of Sec. VI for the evaluation of solutions of the variational equations of Sec. V are our main
original result. We have arrived at a broad framework that presents formidable technical difficulties, which
is not the same as {\it solving} our equations. For this purpose, one should solve completely the following
problems:
\vskip 0.3cm
\noindent
(i) First, how to find periodic solutions of the Hamiltonian equations (4.15) when $\rho=0$. From Eqs. 
(6.27)-(6.31), this means having to solve the infinite system of equations
\begin{equation}
\sum_{l=0}^{\infty}D_{il}\omega_{il}\cos(\omega_{il}t+\varphi_{il})
=\sum_{l=0}^{\infty}E_{il}\sin(\omega_{il}t+\gamma_{il}), \; \forall i=1,...,6,
\label{(7.1)}
\end{equation}
\begin{eqnarray}
\sum_{l=0}^{\infty}E_{il}\omega_{il}\cos(\omega_{il}t+\gamma_{il})
&=& -{A_{0}\over m}\sum_{l=0}^{\infty}D_{il}\sin(\omega_{il}t+ \gamma_{il}) 
\nonumber \\
&+&  {B_{0}\over m}\sum_{l=0}^{\infty}D_{i+3,l}\sin(\omega_{i+3,l}t+ \gamma_{i+3,l}), \;
\forall i=1,2,3,
\label{(7.2)}
\end{eqnarray}
\begin{eqnarray}
\sum_{l=0}^{\infty}E_{il}\omega_{il}\cos(\omega_{il}t+\gamma_{il})
&=& -{C_{0}\over \mu}\sum_{l=0}^{\infty}D_{il}\sin(\omega_{il}t+ \gamma_{il}) 
\nonumber \\
&+&  {B_{0}\over \mu}\sum_{l=0}^{\infty}D_{i-3,l}\sin(\omega_{i-3,l}t+ \gamma_{i-3,l}), \;
\forall i=4,5,6.
\label{(7.3)}
\end{eqnarray}
\vskip 0.3cm
\noindent
(ii) Second, how to solve the variational equations through Eqs. (6.20) and (6.21), or
(6.25) and (6.26), when the matrix $M_{ik}^{0}$ is evaluated along a solution of Eqs. (7.1)-(7.3).
In Refs. \cite{P1890,P1892}, Poincar\'e obtained an algebraic equation of third degree for the
square of $\alpha_{1}$, which was the hardest part of the calculation, but we do not see an 
analogous equation for the square of $\alpha_{1}$ in our case.
\vskip 0.3cm
\noindent
(iii) Third, what is the counterpart, if any, of the variety of periodic and asymptotic solutions 
found by Poincar\'e \cite{P1890,P1892}, i.e., more precisely:
\vskip 0.3cm
\noindent
(iii-a) Periodic solutions of the Hamiltonian equations (4.15) with nonvanishing values of $\rho$,
e.g.
\begin{equation}
x_{l}(t)=\psi_{l}^{0}(t)+(\rho-\rho_{0})^{1 \over 2}\psi_{l}^{(1)}(t)
+(\rho-\rho_{0})\psi_{l}^{(2)}(t)
+(\rho-\rho_{0})^{3 \over 2}\psi_{l}^{(3)}(t)+...,
\label{(7.4)}
\end{equation}
where $\psi_{l}^{0}(t)$ has period $T$, while $\psi_{l}^{(1)}(t),\psi_{l}^{(2)}(t),\psi_{l}^{(3)}(t)$
have period equal to an integer multiple of $T$.
\vskip 0.3cm
\noindent
(iii-b) Asymptotic solutions of Eqs. (4.15) of the first kind, for which
\begin{equation}
x_{i}(t)=\varphi_{i}(t)+A{\rm e}^{-\alpha t}\theta_{i}^{(1)}(t)
+A^{2}{\rm e}^{-2 \alpha t}\theta_{i}^{(2)}(t)
+A^{3}{\rm e}^{-3 \alpha t}\theta_{i}^{(3)}(t)+...,
\label{(7.5)}
\end{equation}
where $\varphi_{i}(t)$ is an unstable periodic solution, $A$ is an arbitrary integration constant,
$\alpha$ is a positive characteristic exponent, $\theta_{i}^{(1)}(t),\theta_{i}^{(2)}(t)...$ have
period $T$. At sufficiently large positive values of $t$ such series are convergent. As
$t \rightarrow \infty$, such solutions approach asymptotically the unstable periodic solution
$\varphi_{i}(t)$.
\vskip 0.3cm
\noindent
(iii-c) Asymptotic solutions of Eqs. (4.15) of the second kind, for which
\begin{equation}
x_{i}(t)=\varphi_{i}(t)+B{\rm e}^{\alpha t}\omega_{i}^{(1)}(t)
+B^{2}{\rm e}^{2 \alpha t}\omega_{i}^{(2)}(t)
+B^{3}{\rm e}^{3 \alpha t}\omega_{i}^{(3)}(t)+...,
\label{(7.6)}
\end{equation}
where $B$ is a new integration constant,
$\alpha$ is again the positive characteristic exponent, and the functions $\omega$ are of
the same functional form as the functions $\theta$ occurring in (7.5).  
At sufficiently large negative values of $t$ such series are convergent. As
$t \rightarrow - \infty$, such solutions approach asymptotically the unstable periodic solution
$\varphi_{i}(t)$.
\vskip 0.3cm
\noindent
(iii-d) Doubly asymptotic solutions which are represented by (7.6) if $t<0$ and $|t|$ is very large,
and by (7.5) if $t>0$ and $|t|$ is very large. The corresponding orbit, which initially differs
slightly from the unstable periodic solution, departs gradually from it at first, and after having
departed significantly from it ends up by approaching asymptotically the unstable periodic solution.
At finite values of $t$, there exist intervals of this time variable where neither (7.5) nor (7.6)
converges in Newtonian physics \cite{P1890,P1892}.

One should notice that the actual evaluation of periodic solutions of the full three-body problem within the
framework of parametrized post-Newtonian formalism is still in its infancy, since, to the best of our knowledge, only 
results for the circular restricted three-body problem are available so far \cite{HuangWu}, unlike the case of
Newtonian theory, where, after centuries of efforts, some periodic solutions of the full three-body problem are
explicitly known by now \cite{Chenciner00}. The years to come will hopefully tell us whether the scheme described
by our Sec. VI may have observational consequences (see also the numerical estimates in Appendix A) 
in orbital motion physics and in the experimental search for quantum gravity effects (see below).

A naturally occurring question is to what extent is it legitimate to keep using the Lagrangian and
Hamiltonian frameworks of classical mechanics, jointly with its set of variational equations, once the
quantum corrections of Refs. \cite{D94,D94c,D03} have been obtained. As far as we can see, a possible answer
is as follows. The work of Refs. \cite{D94,D94c,D03} deals with the leading long distance quantum corrections
to the Newtonian potential, and leads, by construction, to low-energy effects, here considered in the
solar system. These result entirely from the Einstein-Hilbert part of the full Lagrangian of gravity.
The high-energy effects are instead ruled by terms of higher order in the curvature, e.g.
$$
R_{\alpha \beta \gamma \delta}R^{\alpha \beta \gamma \delta}, \;
R_{\alpha \beta}R^{\alpha \beta}, \;
R^{2}, \; \Box R, \;
R_{\alpha \beta}^{\; \; \; \lambda \mu} \;
R_{\lambda \mu}^{\; \; \; \nu \rho} \;
R_{\nu \rho}^{\; \; \; \alpha \beta}
$$
in the quantum effective action (the generating functional of one-particle irreducible diagrams),
which lead however to quantum corrections at long distances severely suppressed with respect to the ones
considered in the papers \cite{D94,D94c,D03}. Of course, our scheme belongs to the family of hybrid schemes
in theoretical physics, discussed in detail in the Introduction of our previous paper \cite{BE14}.

Furthermore, the use of classical mechanics is suggested by the very nature of the quantum corrections
obtained in Refs. \cite{D94,D94c,D03}: a post-Newtonian term, and a quantum term containing a very small
parameter, i.e. the square of Planck length. 

Last, but not least, our predictions have chances of being testable against observations. In the
Earth-Moon-satellite system, we find, with the notation in appendix A, that the planetoid (i.e. satellite)
coordinates at the Lagrangian points of stable equilibrium are
\begin{equation}
x_{Q}=1.8732985853448734 \cdot 10^{8} {\rm m}, \;
y_{Q}=\pm 3.3255375505843085 \cdot 10^{8} {\rm m}, 
\label{(7.7)}
\end{equation}
whereas the classical Newtonian values are
\begin{equation}
x_{C}=1.8732985852568874 \cdot 10^{8} {\rm m}, \;
y_{C}= \pm 3.3255375505322444 \cdot 10^{8} {\rm m}.
\label{(7.8)}
\end{equation}
This means that we predict a quantum correction to the $x$ coordinate given by
\begin{equation}
x_{Q}-x_{C} \approx 0.879 \; {\rm cm},
\label{(7.9)}
\end{equation}
while
\begin{equation}
|y_{Q}|-|y_{C}| \approx 0.52 \; {\rm cm}.
\label{(7.10)}
\end{equation}
Interestingly, these corrections are within reach of current technology, and we arrive at a 
prediction of low-energy quantum gravity effects in the solar system, which was, to our knowledge,
quite unexpected. In the near future we hope to be able to propose measurements aimed at testing
such an effect, which can receive careful consideration, in light of the broad interest of the
scientific community in the applications of Lagrangian points in the solar system
\cite{Prado,Gomez,Simo}.

\acknowledgments 
The authors are indebted to John Donoghue for enlightening correspondence, and to Massimo Cerdonio
and Alberto Vecchiato for conversations. 
G. E. is grateful to the Dipartimento di Fisica of
Federico II University, Naples, for hospitality and support. 

\begin{appendix}

\section{Closer look at stable equilibrium points of the restricted three-body problem}

In Ref. \cite{BE14}, with the notation described therein, according to which the planetoid is
at distance $r$ from the body $A$ of mass $\alpha$ and coordinates $(-a,0)$, and at distance
$s$ from the body $B$ of mass $\beta$ and coordinates $(b,0)$, the coordinates of stable 
equilibrium points of the planetoid are (hereafter $l \equiv (a+b)$)
\begin{equation}
x(l)={(r^{2}(l)-s^{2}(l)+b^{2}-a^{2})\over 2(a+b)},
\label{(A1)}
\end{equation}
\begin{equation}
y_{\pm}(l)=\pm \sqrt{r^{2}(l)-x^{2}(l)-2ax(l)-a^{2}},
\label{(A2)}
\end{equation}
where 
\begin{equation}
r(l)={1 \over w_{+}(l)}, \;
s(l)={1 \over u_{+}(l)},
\label{(A3)}
\end{equation}
$w_{+}$ and $u_{+}$ being the positive roots of the algebraic equation of fifth degree
\begin{equation}
\sum_{k=0}^{5}\zeta_{k}w^{k}=0, \;
\sum_{k=0}^{5}{\tilde \zeta}_{k}u^{k}=0,
\label{(A4)}
\end{equation}
where
\begin{equation}
\zeta_{5}=1, \; \zeta_{4}={2 \over 3}{\kappa_{1}\over \kappa_{2}}
{G(m+\alpha)\over c^{2}l_{P}^{2}}, \;
\zeta_{3}={1 \over 3 \kappa_{2}}{1 \over l_{P}^{2}},
\label{(A5)}
\end{equation}
\begin{equation}
\zeta_{2}=\zeta_{1}=0, \; \zeta_{0}=-{1 \over 3 \kappa_{2}}
{1 \over l_{P}^{2}l^{3}},
\label{(A6)}
\end{equation}
\begin{equation}
{\tilde \zeta}_{k}=\zeta_{k} \; \forall k=0,1,2,3,5, \;
{\tilde \zeta}_{4}={2 \over 3}{\kappa_{3}\over \kappa_{2}}
{G(m+\beta)\over c^{2}l_{P}^{2}}.
\label{(A7)}
\end{equation}
In Ref. \cite{BE14} we have solved numerically such algebraic equations, since no general algorithm
exists for solving algebraic equations of fifth or higher degree. However, since the left-hand side
of Eqs. (A4) is a fairly simple polynomial function, the basic rules for studying functions of a real
variable provide already a valuable information. For example, one has
\begin{equation}
f'(w)=w^{2}(3 \zeta_{3}+4 \zeta_{4}w+5w^{2}),
\label{(A8)}
\end{equation}
which therefore vanishes either at $w=0$ or at
\begin{equation}
w_{1}=-{2 \over 5}\zeta_{4}+{1 \over 5}\sqrt{4 \zeta_{4}^{2}
-15 \zeta_{3}^{2}},
\label{(A9)}
\end{equation}
\begin{equation}
w_{2}=-{2 \over 5}\zeta_{4}-{1 \over 5}\sqrt{4 \zeta_{4}^{2}
-15 \zeta_{3}^{2}}.
\label{(A10)}
\end{equation}
By virtue of (A5), such roots are real provided that
\begin{equation}
{(\kappa_{1})^{2}\over \kappa_{2}} > {45 \over 16}{c^{4}\over G^{2}(m+\alpha)}l_{P}^{2},
\label{(A11)}
\end{equation}
which is satisfied in the Sun-Earth-Moon and Jupiter-Ganimede-Adrastea systems by virtue of the small value
of the Planck length. The roots $w_{1}$ and $w_{2}$ are therefore both negative. Moreover, the second 
derivative of $f$ reads as
\begin{equation}
f''(w)=2w(3 \zeta_{3}+6 \zeta_{4}w +10 w^{2}) \equiv 2w g(w).
\label{(A12)}
\end{equation}
The point $w=0$ is therefore a flex point, while the sign of $f''$ at $w_{1}$ and $w_{2}$, and hence maxima
or minima of $f$, is governed by the sign of the second degree polynomial 
$g(w) \equiv 3 \zeta_{3}+6 \zeta_{4} w +10 w^{2}$.

Interestingly, from Eqs. (A1) and (A2) we find for the Sun-Earth-Moon system the quantum corrected planetoid
(i.e. the Moon) coordinates at equilibrium
\begin{equation}
x_{Q}=7,479978 \cdot 10^{10} {\rm m}, \; y_{Q}=1,29573 \cdot 10^{11} {\rm m},
\label{(A13)}
\end{equation}
to be compared with the classical Newtonian values
\begin{equation}
x_{C}=7,479955 \cdot 10^{10} {\rm m}, \; y_{C}=1,29557 \cdot 10^{11} {\rm m}.
\label{(A14)}
\end{equation}
Moreover, for the Jupiter-Ganimede-Adrastea system, we find the quantum corrected planetoid 
coordinates (i.e. Adrastea) at equilibrium
\begin{equation}
x_{Q}=5,349183 \cdot 10^{8} {\rm m}, \; y_{Q}=9,2698 \cdot 10^{8} {\rm m},
\label{(A15)}
\end{equation}
whereas the classical Newtonian values are
\begin{equation}
x_{C}=5,349167 \cdot 10^{8} {\rm m}, \; y_{C}=9,2665 \cdot 10^{8} {\rm m}.
\label{(A16)}
\end{equation}
As one can see, in both cases, the $x$-values start differing at the fifth decimal digit, while the
$y$-values may start differing at the fourth or third decimal digit.

\section{Definition of characteristic exponents}

Following Refs. \cite{P1890,P1892}, consider the differential
equations
\begin{equation}
{{\rm d}\over {\rm d}t}x_{i}=X_{i},
\label{(B1)}
\end{equation}
and suppose they admit a periodic solution
\begin{equation}
x_{i}=\varphi_{i}(t).
\label{(B2)}
\end{equation}
We can now consider small disturbances of Eqs. (B1) by setting
\begin{equation}
x_{i}=\varphi_{i}(t)+\xi_{i},
\label{(B3)}
\end{equation}
and neglecting the squares of the $\xi_{i}$. We are therefore 
studying the linearized perturbative regime for Eqs. (B1).
This procedure leads to the first-order equations
\begin{equation}
{{\rm d}\over {\rm d}t}\xi_{i}=\sum_{k=1}^{n}
{\partial X_{i}\over \partial x_{k}}\xi_{k},
\label{(B4)}
\end{equation}
known as the variational equations \cite{Pars65}.
These equations are linear with respect to the $\xi_{k}$, and their
coefficients ${\partial X_{i}\over \partial x_{k}}$, where $x_{i}$
should be eventually replaced by $\varphi_{i}(t)$, are periodic
functions of the time variable $t$. Hence we have to integrate
linear differential equations with periodic coefficients. The general
form of the solutions of these equations has been known for centuries;
one obtains $n$ particular solutions of the following form:
\begin{equation}
\xi_{1}={\rm e}^{\alpha_{k}t}S_{1k}, \;
\xi_{2}={\rm e}^{\alpha_{k}t}S_{2k}, \; ..., \;
\xi_{n}={\rm e}^{\alpha_{k}t}S_{nk},
\label{(B5)}
\end{equation}
for all $k=1,2,...,n$, the $\alpha_{k}$ being constants and the
$S_{ik}$ being periodic functions of $t$ with the same period as the
$\varphi_{i}(t)$. The constants $\alpha_{k}$ are said to be
the {\it characteristic exponents} of the periodic solutions
\cite{P1890,P1892}. Our equations of motion (4.15) belong to the
general family expressed by (B1). 

\section{The matrix $M_{ik}^{0}$}

For the matrix of partial derivatives defined in Eq. (6.19) we find (with the understanding
that a subscript like ${}_{,k}$ denotes partial derivative with respect to $x_{k}$, for all
$k=1,...,6$)
\begin{equation}
M_{11}^{0}={1 \over m}(x_{1}A_{0,1}+A_{0}-x_{4}B_{0,1}),
\label{(C1)}
\end{equation}
\begin{equation}
M_{12}^{0}={1 \over m}(x_{1}A_{0,2}-x_{4}B_{0,2}),
\label{(C2)}
\end{equation}
\begin{equation}
M_{13}^{0}={1 \over m}(x_{1}A_{0,3}-x_{4}B_{0,3}),
\label{(C3)}
\end{equation}
\begin{equation}
M_{14}^{0}={1 \over m}(x_{1}A_{0,4}-x_{4}B_{0,4}-B_{0}),
\label{(C4)}
\end{equation}
\begin{equation}
M_{15}^{0}={1 \over m}(x_{1}A_{0,5}-x_{4}B_{0,5}),
\label{(C5)}
\end{equation}
\begin{equation}
M_{16}^{0}={1 \over m}(x_{1}A_{0,6}-x_{4}B_{0,6}),
\label{(C6)}
\end{equation}
\begin{equation}
M_{21}^{0}={1 \over m}(x_{2}A_{0,1}-x_{5}B_{0,1}),
\label{(C7)}
\end{equation}
\begin{equation}
M_{22}^{0}={1 \over m}(x_{2}A_{0,2}+A_{0}-x_{5}B_{0,2}),
\label{(C8)}
\end{equation}
\begin{equation}
M_{23}^{0}={1 \over m}(x_{2}A_{0,3}-x_{5}B_{0,3}),
\label{(C9)}
\end{equation}
\begin{equation}
M_{24}^{0}={1 \over m}(x_{2}A_{0,4}-x_{5}B_{0,4}),
\label{(C10)}
\end{equation}
\begin{equation}
M_{25}^{0}={1 \over m}(x_{2}A_{0,5}-x_{5}B_{0,5}-B_{0}),
\label{(C11)}
\end{equation}
\begin{equation}
M_{26}^{0}={1 \over m}(x_{2}A_{0,6}-x_{5}B_{0,6}),
\label{(C12)}
\end{equation}
\begin{equation}
M_{31}^{0}={1 \over m}(x_{3}A_{0,1}-x_{6}B_{0,1}),
\label{(C13)}
\end{equation}
\begin{equation}
M_{32}^{0}={1 \over m}(x_{3}A_{0,2}-x_{6}B_{0,2}),
\label{(C14)}
\end{equation}
\begin{equation}
M_{33}^{0}={1 \over m}(x_{3}A_{0,3}+A_{0}-x_{6}B_{0,3}),
\label{(C15)}
\end{equation}
\begin{equation}
M_{34}^{0}={1 \over m}(x_{3}A_{0,4}-x_{6}B_{0,4}),
\label{(C16)}
\end{equation}
\begin{equation}
M_{35}^{0}={1 \over m}(x_{3}A_{0,5}-x_{6}B_{0,5}),
\label{(C17)}
\end{equation}
\begin{equation}
M_{36}^{0}={1 \over m}(x_{3}A_{0,6}-x_{6}B_{0,6}-B_{0}),
\label{(C18)}
\end{equation}
\begin{equation}
M_{41}^{0}={1 \over \mu}(x_{4}C_{0,1}-x_{1}B_{0,1}-B_{0}),
\label{(C19)}
\end{equation}
\begin{equation}
M_{42}^{0}={1 \over \mu}(x_{4}C_{0,2}-x_{1}B_{0,2}),
\label{(C20)}
\end{equation}
\begin{equation}
M_{43}^{0}={1 \over \mu}(x_{4}C_{0,3}-x_{1}B_{0,3}),
\label{(C21)}
\end{equation}
\begin{equation}
M_{44}^{0}={1 \over \mu}(x_{4}C_{0,4}+C_{0}-x_{1}B_{0,4}),
\label{(C22)}
\end{equation}
\begin{equation}
M_{45}^{0}={1 \over \mu}(x_{4}C_{0,5}-x_{1}B_{0,5}),
\label{(C23)}
\end{equation}
\begin{equation}
M_{46}^{0}={1 \over \mu}(x_{4}C_{0,6}-x_{1}B_{0,6}),
\label{(C24)}
\end{equation}
\begin{equation}
M_{51}^{0}={1 \over \mu}(x_{5}C_{0,1}-x_{2}B_{0,1}),
\label{(C25)}
\end{equation}
\begin{equation}
M_{52}^{0}={1 \over \mu}(x_{5}C_{0,2}-x_{2}B_{0,2}-B_{0}),
\label{(C26)}
\end{equation}
\begin{equation}
M_{53}^{0}={1 \over \mu}(x_{5}C_{0,3}-x_{2}B_{0,3}),
\label{(C27)}
\end{equation}
\begin{equation}
M_{54}^{0}={1 \over \mu}(x_{5}C_{0,4}-x_{2}B_{0,4}),
\label{(C28)}
\end{equation}
\begin{equation}
M_{55}^{0}={1 \over \mu}(x_{5}C_{0,5}+C_{0}-x_{2}B_{0,5}),
\label{(C29)}
\end{equation}
\begin{equation}
M_{56}^{0}={1 \over \mu}(x_{5}C_{0,6}-x_{2}B_{0,6}),
\label{(C30)}
\end{equation}
\begin{equation}
M_{61}^{0}={1 \over \mu}(x_{6}C_{0,1}-x_{3}B_{0,1}),
\label{(C31)}
\end{equation}
\begin{equation}
M_{62}^{0}={1 \over \mu}(x_{6}C_{0,2}-x_{3}B_{0,2}),
\label{(C32)}
\end{equation}
\begin{equation}
M_{63}^{0}={1 \over \mu}(x_{6}C_{0,3}-x_{3}B_{0,3}-B_{0}),
\label{(C33)}
\end{equation}
\begin{equation}
M_{64}^{0}={1 \over \mu}(x_{6}C_{0,4}-x_{3}B_{0,4}),
\label{(C34)}
\end{equation}
\begin{equation}
M_{65}^{0}={1 \over \mu}(x_{6}C_{0,5}-x_{3}B_{0,5}),
\label{(C35)}
\end{equation}
\begin{equation}
M_{66}^{0}={1 \over \mu}(x_{6}C_{0,6}+C_{0}-x_{3}B_{0,6}).
\label{(C36)}
\end{equation}

\end{appendix}

\end{document}